\documentclass[10pt,journal,compsoc]{IEEEtran}
\usepackage{mystyle}
\DeclareAcronym{api}{
	short = API,
	long = {Application Program Interface}
}

\DeclareAcronym{awgn}{
	short = AWGN,
	long = {additive white Gaussian noise}
}

\DeclareAcronym{vae}{
	short = VAE,
	long = {Variational AutoEncoder}
}

\DeclareAcronym{bert}{
	short = BERT,
	long = {Bidirectional Encoder Representations from Transformers}
}

\DeclareAcronym{roberta}{
	short = RoBERTa,
	long = {Robustly optimized BERT approach}
}

\DeclareAcronym{ast}{
	short = AST,
	long = {Abstract Syntax Tree}
}

\DeclareAcronym{bpe}{
	short = BPE,
	long = {Byte-Pair Encoding}
}

\DeclareAcronym{cfg}{
	short = CFG,
	long = {Control Flow Graph}
}

\DeclareAcronym{dcg}{
	short = DCG,
	long = {Discounted Cumulative Gain}
}

\DeclareAcronym{gpt}{
	short = GPT,
	long = {Generative Pretrained Transformer}
}

\DeclareAcronym{ir}{
	short = IR,
	long = {Information Retrieval}
}

\DeclareAcronym{lstm}{
	short = LSTM,
	long = {Long Short-Term Memory}
}

\DeclareAcronym{clm}{
	short = CLM,
	long = {Casual Language Modeling}
}

\DeclareAcronym{mlm}{
	short = MLM,
	long = {Masked Language Modeling}
}

\DeclareAcronym{mem}{
	short = MEM,
	long = {Multimodal Embedding Model}
}

\DeclareAcronym{cp}{
	short = CP,
	long = {Continuous Pretraining}
}

\DeclareAcronym{if}{
	short = IF,
	long = {Intermediate Finetuning}
}

\DeclareAcronym{mmpf}{
	short = MMPF,
	long = {Massive Multitask Pre-Finetuning}
}

\DeclareAcronym{aif}{
	short = AIF,
	long = {Adaptive Intermediate Finetuning}
}

\DeclareAcronym{mrr}{
	short = MRR,
	long = {Mean Reciprocal Rank}
}

\DeclareAcronym{ndcg}{
	short = NDCG,
	long = {Normalized Discounted Cumulative Gain}
}

\DeclareAcronym{nlp}{
	short = NLP,
	long = {Natural Language Processing}
}

\DeclareAcronym{nlp_pt}{
	short = NLP\textsubscript{PT},
	long = {Next Line Prediction}
}

\DeclareAcronym{nmt}{
	short = NMT,
	long = {Neural Machine Translation}
}

\DeclareAcronym{nsp}{
	short = NSP,
	long = {Next Sentence Prediction}
}

\DeclareAcronym{rnn}{
	short = RNN,
	long = {Recurrent Neural Network}
}

\DeclareAcronym{cnn}{
	short = CNN,
	long = {Convolutional Neural Network}
}

\DeclareAcronym{tf-idf}{
	short = tf-idf,
	long = {term frequency–-inverse document frequency}
}

\DeclareAcronym{anova}{
	short = ANOVA,
	long = {ANalysis Of VAriance}
}

\DeclareAcronym{da}{
	short = DA,
	long = {Domain-Adaptive}
}

\DeclareAcronym{ta}{
	short = TA,
	long = {Task-Adaptive}
}

\DeclareAcronym{ma}{
	short = MA,
	long = {Multiphase Adaptive}
}

\DeclareAcronym{ca}{
	short = CA,
	long = {Concept Annotation}
}

\DeclareAcronym{ce}{
	short = CE,
	long = {Concept Extrapolation}
}

\DeclareAcronym{ci}{
	short = CI,
	long = {Concept Interpolation}
}

\DeclareAcronym{gru}{
	short = GRU,
	long = {Gated Recurrent Unit}
}

\DeclareAcronym{sota}{
	short = SOTA,
	long = {state-of-the-art}
}

\DeclareAcronym{lcs}{
	short = LCS,
	long = {Longest Common Sequences}
}

\begin{document}

\newcommand{\myauthornote}[3]{{\color{#2} {\sc #1}: #3}}
\newcommand{\hy}[1]{\myauthornote{HY}{red}{#1}}
\newcommand{\cy}[1]{\textcolor{blue}{\textbf{Chen}: #1}}
\newcommand{\gu}[1]{\myauthornote{Gu}{blue}{#1}}

\title{Neuron Patching: Semantic-based Neuron-level\\Language Model Repair for Code Generation}
\author{Jian~Gu,
~Aldeida~Aleti,
~Chunyang~Chen,
and~Hongyu~Zhang
\IEEEcompsocitemizethanks{
\IEEEcompsocthanksitem Jian Gu, Aldeida Aleti are with Monash University, Australia.
\protect\\E-mail: \{jian.gu, aldeida.aleti\}@monash.edu
\IEEEcompsocthanksitem Chunyang Chen is with Technical University of Munich, Germany.
\protect\\E-mail: chun-yang.chen@tum.de
\IEEEcompsocthanksitem Hongyu Zhang is with Chongqing University, China.
\protect\\E-mail: hyzhang@cqu.edu.cn
}
}




\IEEEtitleabstractindextext{
\begin{abstract}
Language Models (LMs) have become widely used in software engineering, especially for tasks such as code generation, where they are referred to as code LMs.
These models have proven effective in generating code, making it easier for developers to automate coding activities.
However, research has highlighted a significant limitation: despite their effectiveness, LMs often produce code that is incorrect, buggy, or not fully functional.
Updating these models with limited data can be prohibitively challenging, yet it is essential to maximize their utility. This may require hot-fix techniques (updating models with limited data) to resolve.
In this paper, we propose \ul{M}odel \ul{I}mprovement via \ul{N}euron \ul{T}argeting (\textsc{MINT}), a novel approach for repairing code LMs.
MINT leverages the semantic property of language models to perform neuron-level repairs in a novel way.
Further, by analyzing the relationships between the model's latent representations, the incorrect outputs, and the desired outputs, \textsc{MINT} determines which neurons are worth updating.
This approach ensures that only the neurons crucial to the model's failure are targeted, avoiding unnecessary changes and allowing for a more efficient and precise repair process.
\textsc{MINT} is effective, efficient, and reliable, capable of correcting a neural model by patching a minimum number of neurons (usually one or two neurons).
We introduce new measures to evaluate its generalisability and develop a new benchmark which is made available for further study.
Our approach is evaluated on three coding tasks: line-level code generation, shellcode generation, and intent-to-bash translation. The experimental results demonstrate that the proposed approach significantly outperforms the state-of-the-art in both effectiveness and efficiency measures. With respect to the ExactMatch score, \textsc{MINT} achieves $5.7\%-20.8\%$ improvements in \textsc{StarCoder2-3B}, and $3.9\%-18.5\%$ improvements in \textsc{CodeLlama-7B} concerning the state-of-the-art.
Regarding efficiency, \textsc{MINT} is $32.3\%-74.8\%$ faster than the state-of-the-art.
In addition, we analyze and discuss the side effects of model repair techniques, including the balance between generalization and specificity, and the performance after multiple repairs in succession.
\end{abstract}

}



\maketitle

\section{Introduction}
\label{sec:introduction}
\IEEEPARstart{W}{ith} the rapid development in deep learning, language models (LMs) have shown great potential in many areas, such as computational linguistics and computer vision~\cite{Gao2021MakingPL,Wang2023VisionLLMLL}.
By refining LMs on code-related data~\cite{Le2020DeepLF}, researchers have shown that these models, also known as \textit{Code LMs}, have great potential in code modeling and generation tasks, such as automated program repair~\cite{Xia2023AutomatedPR}, automated code review~\cite{Tufano2021TowardsAC} and assisted programming~\cite{Chen2021EvaluatingLL}. Recent studies have also extended the application of code LMs to programming education and competitions~\cite{Zhang2022RepairingBI,Li2022CompetitionlevelCG}.

Over time, LMs can become outdated or pose safety risks, regularly updating models with limited data is important to realize their value~\cite{Dong2022CalibratingFK}.
For example, they may require updates to handle breaking changes of dependencies~\cite{Jayasuriya2023UnderstandingBC}, or major revisions of popular frameworks (e.g., React 18).
Besides, LMs constantly producing programs with bugs or vulnerabilities can be a nightmare~\cite{Li2024BadEditBL,Li2024BackdoorLLMAC}.
Rather than relying on extensive post-processing efforts, such as fixing vulnerabilities in the generated code, a better practice is to directly repair the source, namely the models.
Additionally, language models accidentally expose failures that require flexible corrections, such as wrongly stating $9.8 < 9.11$~\cite{web_transluce_20241031}.
The failures, though seemingly small, can accumulate and lead to significant consequences in downstream tasks.


Existing techniques, including model finetuning and rule-based methods~\cite{Hu2021LoRALA,Gu2022AssembleFM,Dziri2021NeuralPH}, are not the solution due to several reasons~\cite{Yao2023EditingLL}.
For instance, an LM-based product or an LM-driven system may require a hotfix (or efficient patch), making it impractical to finetune the model; and costly for the required computational resources and data resources.
However, finetuning with a small amount of data is likely to cause overfitting and catastrophic forgetting~\cite{Kirkpatrick2016OvercomingCF}.
Meanwhile, rule-based methods fail due to the lack of flexibility.
LMs leverage high-dimensional latent representations to hold vast information, allowing corrections to the information will automatically enhance various related model behaviors due to their interconnected nature. In contrast, rule-based methods rely on manually defined rules that become increasingly complex and hard to manage as accumulated changes grow, making them cumbersome and overly complex.

To solve failures of language models, the countermeasure is \textit{language model repair}.
Model repair alters the parameters of a neural model, transforming its original state into a new altered state.
The research gap is, \emph{there are no methods for repairing language models, especially for general next-token prediction}.
This paper explores how to effectively, efficiently, and reliably update language models to fix their incorrect behaviors (LM failures).
When language models display unexpected behaviors, \ul{in-place and targeted interventions based on limited data} act as hot patches to repair model failures.
Considering the fundamental role that LMs will play in automating various coding tasks~\cite{Bommasani2021OnTO}, addressing the limitations of code LMs will produce important benefits for software developers or agents~\cite{Xia2024AgentlessDL}.

We propose a semantic-based neuron-level approach for repairing language models: \ul{M}odel \ul{I}mprovement via \ul{N}euron \ul{T}argeting, short for \textbf{\textsc{MINT}}. MINT is a novel model repair technique for (code) language models.
Its mechanism is composed of three steps: i) attributing scores to neurons based on their contribution to model failures; ii) estimating the semantic-based patches for updating neuron parameters; iii) locating neurons that can solve the model failures and applying the patches.
The novelty of our approach lies in leveraging the semantic property of language models to enable neuron-level model repair, including:
(1) It enables the updating of fewer parameters, helping to preserve the original model behaviors while allowing for adaptation~\cite{Peters2019ToTO,He2021TowardsAU}. This approach reduces catastrophic forgetting and side effects, by repairing the model at the granularity of individual neurons.
(2) Solving failures is converted as reducing the semantic difference between two specific latent representations, so the neuron patches can be estimated quicker without iterative attempts.
Further, (3) we distinguish the neurons solving model failures from the neurons causing model failures, and repair models by patching the former.
This is different from program repair where the line of code causing failures is the line of code solving failures.

\textsc{MINT} is fast and effective in repairing models. It can solve model failures with few exemplary data, by updating one or two neurons, without causing side effects.
In our experiments, we studied its effectiveness and efficiency in three coding tasks: line-level code generation, shellcode generation, and intent-to-bash translation.
The results show that \textsc{MINT} significantly outperforms state-of-the-art (SOTA) approaches in terms of effectiveness and efficiency.
In addition, we investigate the side effects of the different variants of our approach by measuring key attributes of model repair: generalization and specificity.


\smallskip
\noindent
To summarize, the contributions of this paper are as follows:
\begin{itemize}
    \item (Task) To the best of our knowledge, this is the first work focused on repairing language models for high-quality code generation.
    \item (Approach) We introduce a novel model repair approach to solve model failures in next-token prediction. Our approach is effective, efficient, and shows reliable generalization and specificity;
    \item (Dataset) We build a new benchmark that allows the evaluation of generalization and specificity of model repair for code models. The benchmark and the replication package are available online for further research \footnote{\url{https://anonymous.4open.science/r/mint-A35E}}.
\end{itemize}


\section{Background}
\label{sec:background}

\subsection{Structure of Language Models}

Language models are composed of layers of Self-Attention Networks (SAN) and Feed-Forward Networks (FFN), with an embedding matrix at the input and an LM-head matrix at the output, following the Transformer structure~\cite{Vaswani2017AttentionIA}.
% ...
SAN enables the model to assign varying attention weights to tokens in an input sequence, capturing relationships between tokens regardless of their position.
The output from SAN is processed by FFN, which are fully connected layers that refine the information for enhanced feature representation.
% ...
The embedding matrix transforms tokens into dense vectors. These vectors, learned during training, encode the semantic properties of tokens.
% ...
The LM-head matrix converts the latent representations into a probability distribution over the vocabulary, predicting the likelihood of the next token.




\subsection{Language Model Repair for Next-Token Prediction}

Model repair is the process of modifying the parameters of a pretrained neural model $M$ in place, resulting in its original state changing to a new state $M'$.
Assuming a running model has exhibited unexpected behaviors, in-place and targeted treatments can repair the model as hot patches, even with limited data.
This means that instead of performing a full system update, specific corrections are made directly within the model while it is still running.

\begin{figure}[htbp]
    \centering
    \includegraphics[width=1.0\linewidth]{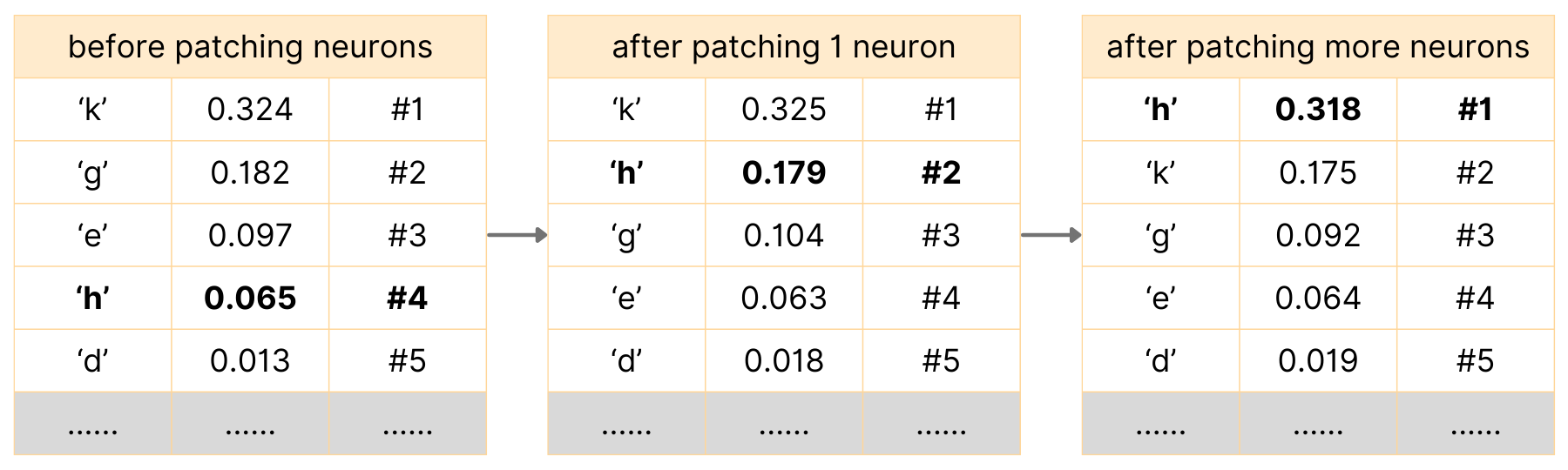}
    \caption{Effects of model repair on vocabulary probabilities.}
    \label{fig:probability}
\end{figure}

In next-token prediction, language models predict the probabilities of outputting each token of the LM vocabulary.
Model repair makes changes to the predicted probabilities.
Let us demonstrate the effects of model repair (and how MINT works), assuming the LM vocabulary is the alphabet, and the target token is `h'. The sorted probabilities on the vocabulary are shown in \cref{fig:probability}.
Initially, the target token `h' is ranked 4th, while the argmax token is `k'. After patching a neuron, the rank of `h' is promoted to the 2nd place. The process of neuron-patching continues until `h' is ranked 1st, as it is the desired target token.

\subsection{Semantics Property of LM Latent Space}

In our previous work, we introduced the vocabulary-defined semantics theory for language models~\cite{Gu2024VocabularyDefinedSL}. The core idea is to associate specific semantic representations in the LM's latent space with each word or token in its vocabulary. This allows us to better understand the semantic meaning of latent representations. To achieve this, we proposed defining a set of special representations that align with vocabulary tokens, leveraging the local isotropy of the semantics property in LM latent space (that is, the semantics tend to have identical or similar values in all directions)~\cite{Cai2021IsotropyIT}.

\begin{figure}[!htb]
    \centering
    \includegraphics[width=0.8\linewidth]{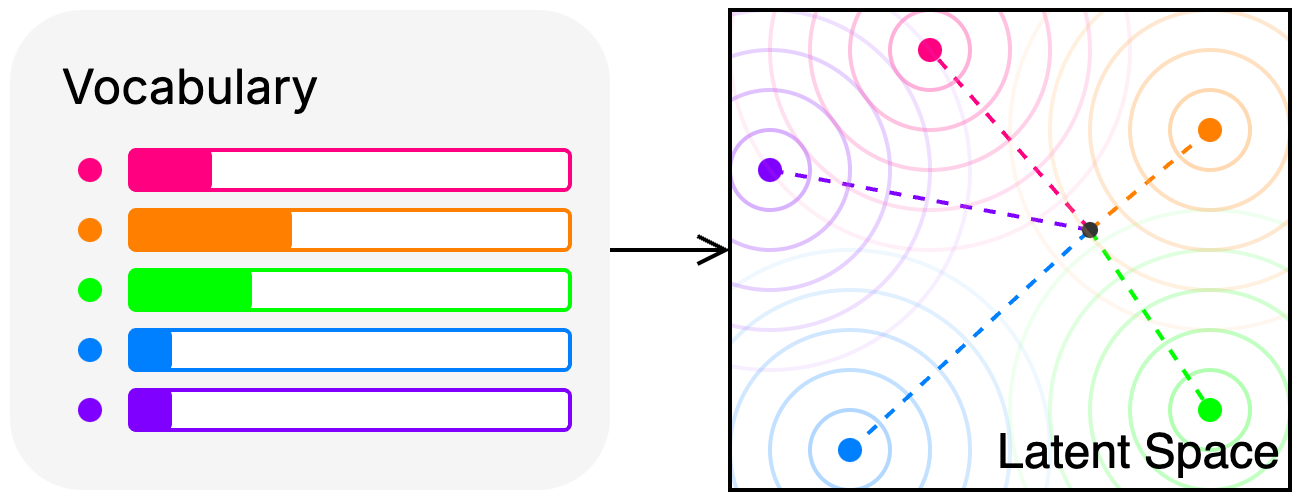}
    \caption{Semantic association of vocabulary and latent space.}
    \label{fig:semantic}
\end{figure}

For each token on the LM vocabulary, there is an associated representation in the LM latent space, termed as ``semantic basis''. These semantic bases serve as reference points that represent the meanings of the tokens in the vocabulary.~\cite{Gu2024VocabularyDefinedSL}.
Furthermore, the semantics of an arbitrary representation in the latent space, represented by the black dot, can be assessed by comparing it to the semantic bases, represented by the colored dots, as shown in \cref{fig:semantic}. The cosine similarities between the latent representation (dark dot) and the semantic bases (colored dots) determine its semantic meaning, which are numerically equivalent to the logits and indicate the probability distribution over the vocabulary. As illustrated in the figure, the shorter colored dashed lines in latent space mean smaller distances, namely bigger similarities to the semantic bases, corresponding to the larger probabilities on the vocabulary.
The ripple patterns of semantic bases indicate the isotropy.

\paragraph{Computation of Semantic Bases}
Both the embedding matrix and LM-head matrix interact with LM vocabulary, we can thereby compute semantic bases in the latent spaces at both the input side and output side.
At the input side, we multiply the onehot embedding of a given token $\vec{e}_{i}$ by the embedding matrix $\mathbb{W}_{i}$ to obtain semantic basis $\vec{s}_{i}$ in the latent space of LM embedding-layer: $\vec{s}_{o}=\vec{e}_{o}\cdot\mathbb{W}_{o}^+$;
At the output side, due to the opposite operation direction between latent space and LM vocabulary, we first compute with the pseudoinverse of LM-head matrix. Then we multiply a given onehot distribution $\vec{e}_{o}$ by the pseudoinverse matrix $\mathbb{W}_{o}^+$ to obtain semantic basis $\vec{s}_{o}$ in the latent space of LM last-layer: $\vec{s}_{i}=\vec{e}_{i}\cdot\mathbb{W}_{i}$.



\section{Approach}
\label{sec:approach}




\textsc{MINT}, short for \ul{M}odel \ul{I}mprovement via \ul{N}euron \ul{T}argeting
\textsc{MINT} is a novel model repair technique for code LMs. The main steps of our approach are described in \cref{algo:repair}.
In the context of next-token prediction for a given LM, including tasks like code generation by code LMs, the correct answer is referred to as the \textit{target token}, while the token with the highest probability is called the \textit{argmax token}.
The two tokens should match; if they do not, it indicates a model failure. In such cases, MINT applies neuron-patching to repair the model.
To solve an LM failure, MINT keeps patching a buggy neuron until either (1) the argmax token becomes the target token, or (2) the number of patched neurons reaches a given quota.
For the former, the LM failure is successfully solved, and the patches to the LM are confirmed.
For the latter, the LM failure skips repairing and the patched neurons are restored with their initial parameters.
In solving an actual model failure, LM tends to be repaired multiple times in succession, where each repair is for one wrongly predicted token.

\begin{algorithm}
\SetKwData{n}{$n$}
\SetKwData{p}{$p$}
\SetKwData{p'}{$p'$}
\SetKwData{cost}{$cost$}
\SetKwData{index}{$index$}
\SetKwData{token}{$token$}
\KwData{prompt $p$; sequence $T$ of $n$ ground truth tokens; $quota$ of the repair cost; the vanilla LM $M$}
\KwResult{token sequence $B$, $D$, $A$ before/during/after model repair; the repaired LM $M'$}

$ B \leftarrow M.\mathrm{predict\_tokens(\p, \n)} $\;

$ p', M' \leftarrow copy(p), copy(M) $\;
\For{$\index \leftarrow 0$ \KwTo $n-1$}{
    \cost $\leftarrow$ 0\;
    $ handle \leftarrow \mathrm{backup\_model}(M') $\;
    \Repeat{$\cost >= quota$}{
        \token $\leftarrow$ $M'.\mathrm{predict\_tokens(\p', 1)}$\;
        \If{$\token == T[\index]$}{
            $ \mathrm{break} $\;
        }
        $ M' \leftarrow \mathrm{neuron\_patching}(M') $\;
        \cost $\leftarrow$ \cost + 1\;
    }
    \tcp{skip repairing}
    \If{$\cost >= quota$}{
        $ M' \leftarrow \mathrm{restore\_model}(handle) $\;
        \token $\leftarrow$ $M'.\mathrm{predict\_tokens(\p', 1)}$\;
    }
    $ p' \leftarrow p' + T[\index] $\;
    $ D[\index] \leftarrow token $\;
}

$ A \leftarrow M'.\mathrm{predict\_tokens(\p, \n)} $\;

\caption{Neuron Patching for LM Repair.}
\label{algo:repair}
\end{algorithm}



\begin{figure}[!ht]
    \centering
    \includegraphics[width=1.0\linewidth]{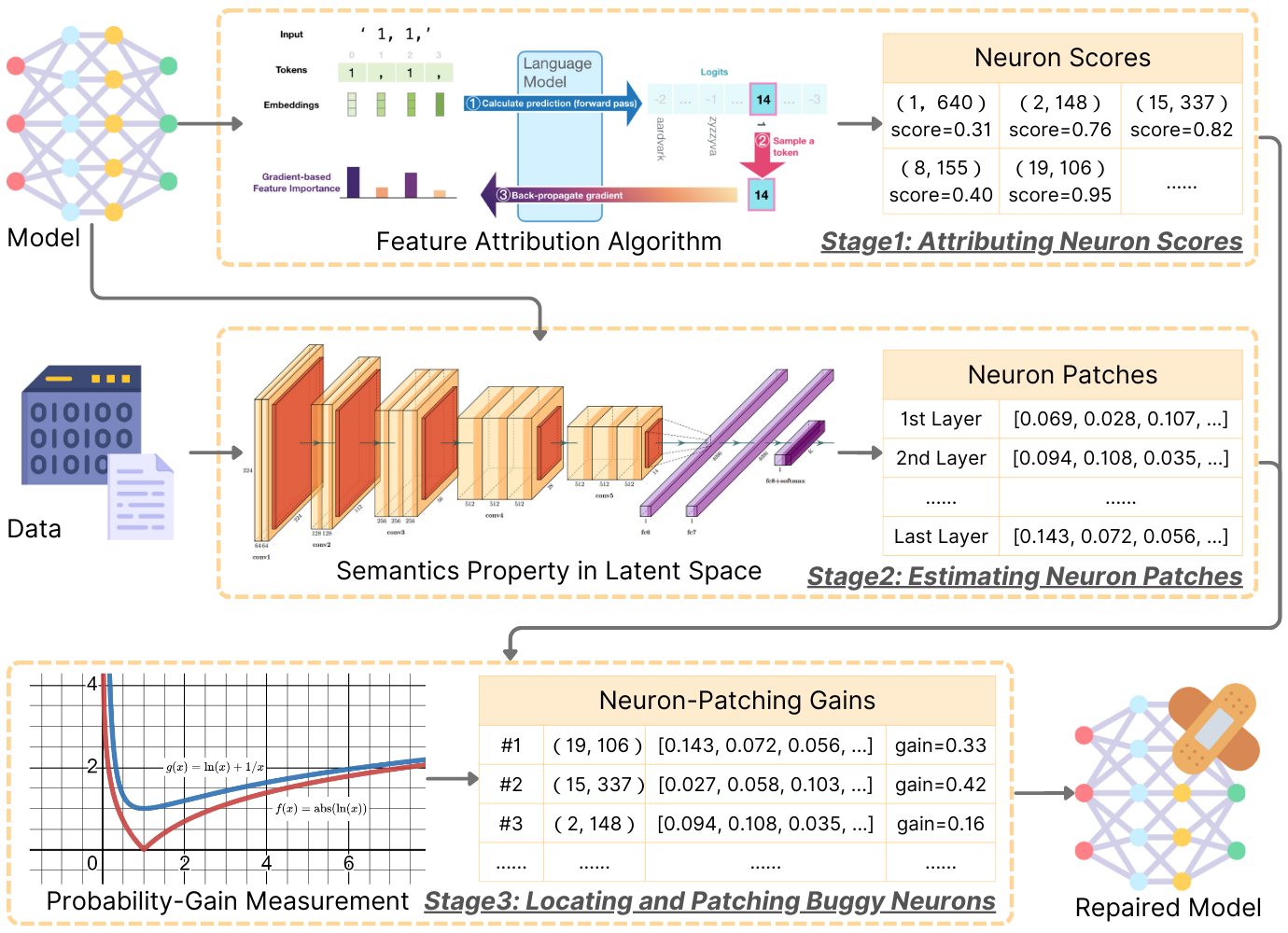}
    \caption{Workflow of our approach for model repair.}
    \label{fig:workflow}
\end{figure}

As illustrated in \cref{fig:workflow}, MINT consists of three stages.
First, MINT applies feature attribution to compute a score for each neuron, indicating the contribution of each neuron to model outputs. The scores reveal the responsibility of each neuron in causing model failures.
Then, we compute the additive parameters in each model layer (corresponding to an individual latent space) leveraging the semantics property in LM latent space. MINT also estimates an adaptive coefficient to adjust the additive parameters in patching each individual neuron.
At last, we measure a gain for each neuron in a simulation to decide the priorities of neurons to patch. We will locate and patch buggy neurons, i.e., neurons that are critical in solving the model failure.
We will describe each stage in detail in the following subsections.

\subsection{Attributing Neuron Scores}


Previous studies have demonstrated that certain neurons in language models have a greater impact than others on the probability of a given output~\cite{Dhamdhere2018HowII,Deng2023UnderstandingAU}.
The contribution of neurons to the model's outputs can be quantified using feature attribution algorithms~\cite{Chattopadhyay2019NeuralNA}, by analyzing activations and gradients~\cite{Zhou2021DoFA}. 
% ...
Activations are the output values of neurons as they process input data,
% ...
while gradients indicate how changes in a neuron's output affect the model's outputs.
% ...

\begin{figure}[htbp]
    \centering
    \includegraphics[width=1.0\linewidth]{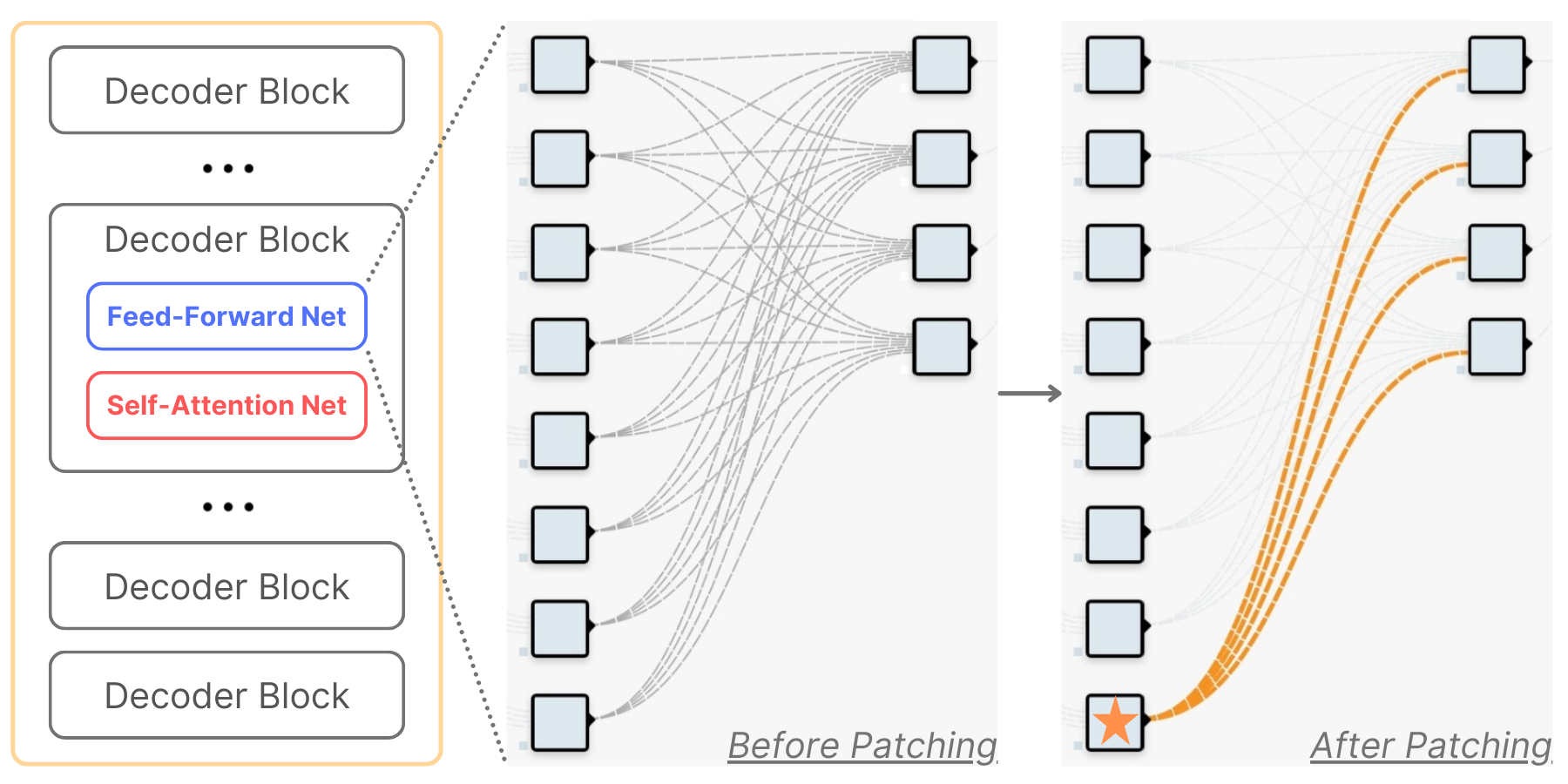}
    \caption{Process of neuron-level model repair in LMs.}
    \label{fig:process}
\end{figure}

When repairing LMs, MINT focuses on the FFN layers because they are integral to storing and processing the learned information~\cite{Geva2020TransformerFL}.
In addition, neurons within the FFN (Feed-Forward Network) hidden layers are referred to as \textit{knowledge neurons}~\cite{Dai2021KnowledgeNI}, meaning they retain essential information of models.
If a neuron in the FFN hidden layer is identified as critical to a model's failure, then it is likely to be patched.
As shown in \cref{fig:process}, the neuron selected for patching is marked with an orange star. The parameters of this neuron, highlighted with orange lines, will be updated during the repair process.

To identify the neurons that contribute the most to incorrect outputs, we employ the feature attribution algorithm \textit{Input X Gradient}~\cite{Ancona2017AUV}. The approach computes an attribution score for each neuron based on its activation signal and gradient signal. A neuron is assigned a higher attribution score if it contributes more than other neurons to the current argmax token. For a given token $t$, \textit{Input X Gradient} multiplies its gradient with the input embedding and then takes the L2 normalization of the resulting vector~\cite{Ancona2017AUV}.
The computation of the attribution score is as follows:

\begin{equation}
\label{eq:input_x_gradient}
\tt{score} = \left\|\nabla_{X_i} f_t\left(X_{1: n}\right) X_i\right\|_2\\
\end{equation}
\begin{align*}
\text{where }
&\text{$X_i$ is the input embedding at step i,}\\
&\text{$\nabla_{X_i} f_t\left(X_{1: n}\right)$ is the gradient of token $t$.}
\end{align*}

The attribution score is an approximated quantization of the contribution of each neuron to the model's outputs.
The neurons are then ranked based on the attribution score and the top-ranked neurons are deemed as the most critical neurons in causing an LM failure.

In the evaluation, we explore variants of the feature attribution algorithm, including: 1) using the activation values of neurons, namely $\tt{score} = \left\|X_i\right\|_2$, labeled \expttag{[attr-actv]}; 2) using randomly generated values, namely $\tt{score} = \mathtt{random}()$, labeled \expttag{[attr-rand]}.

\subsection{Estimating Neuron Patches}

\begin{figure*}[!ht]
    \centering
    \includegraphics[width=1.0\linewidth]{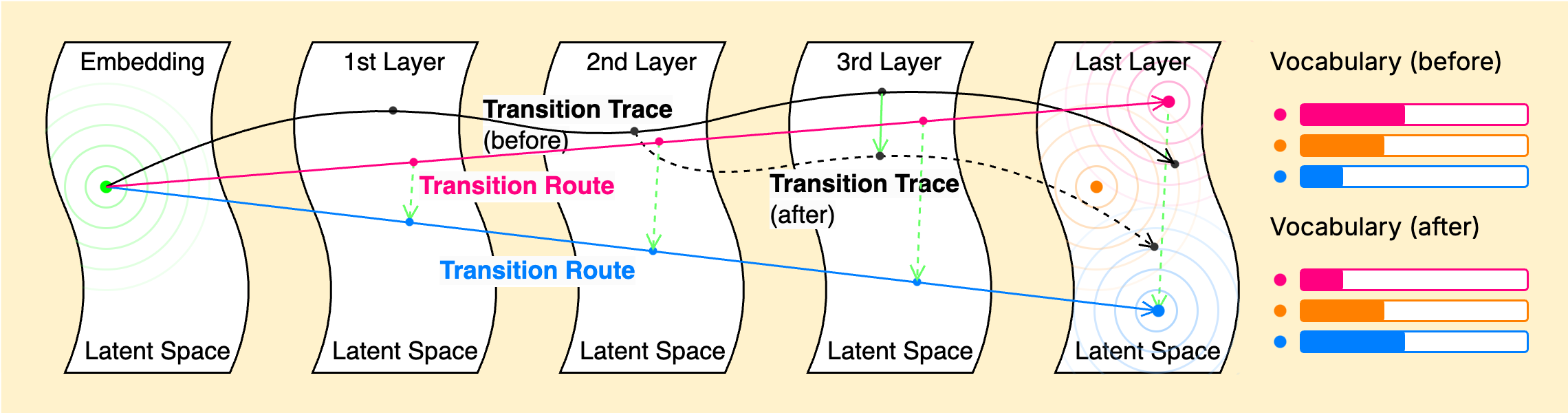}
    \caption{Semantic-based perspective for the mechanism of neuron-patching, associating the latent space and LM vocabulary.}
    \label{fig:perspective}
\end{figure*}

Our approach proposes steering the latent representation of the medium token towards the ground truth by utilizing semantic bases~\cite{Gu2024VocabularyDefinedSL}.
In next-token prediction, the medium token means the last token of the inputs and its latent representation directly decides the output token~\cite{Gu2024ASL}.
%
In neuron-patching, the parameters of buggy neuron are updated in a specific direction using additive parameters applied in a single step, instead of through iterative attempts in arbitrary directions.
The mechanism is grounded in a semantic-based analysis, as illustrated in \cref{fig:perspective}.
The steps of steering the representations in latent space are explained below.



\paragraph{Clarifying the semantic transition in LM latent space}
The latent representation of the medium token undergoes a gradual transition starting from the input-side semantic basis. This transition is referred to as a "transition trace", visualized as the dark solid curve.
We plan virtual transitions which end at output-side semantic bases. We refer to each as a ``transition route'', depicted as the colored solid lines.
In \cref{fig:perspective}, the dark solid curve is close to the red solid line, so that the dark dot in the last-layer latent space is near the red semantic basis, which corresponds to the red token in the vocabulary. However, the ground truth is the blue token, so the latent representation needs to be steered towards the blue semantic basis, ensuring the target token becomes the argmax token.
Essentially, the transition trace (or route) determines how the initial input representation $\vec{s}_{i}$ evolves as it progresses through the model layers, ultimately reaching the output representation, namely $\vec{s}_{o\tt{(argmax)}}$ (or $\vec{s}_{o\tt{(target)}}$).


\paragraph{Computing the semantic difference in LM latent space}
To conduct LM repair, we need the difference between red dots and blue dots.
The red and blue dots in the middle layers represent the virtual intermediate states when following different transition routes, termed as ``semantic anchors'', and their values are unknown.
In our approach, we focus solely on the differences between semantic anchors, termed as ``semantic difference''. This semantic difference indicates the difference between the probabilities of the target token and the argmax token.
% ...
For an LM with $m$ layers, the semantic difference at the $k$-th layer is noted as $\vec{s}_{\tt{\Delta}(k)}$ (where $k \in [0, m] \cap \mathbb{Z}$, and the $0$-th layer indicates the embedding layer), illustrated by the green dashed lines.
% ...
The intuition of MINT is that, semantic anchors following different transition routes tend to exhibit similar patterns of change. 
Based on this idea, the semantic differences in the middle layers can be estimated with semantic bases.
The formula of computing semantic differences tends to be an unknown scaling function, which fulfills the semantic transition from the initial input-side semantic basis $\vec{s}_{i}$ (where $\vec{s}_{\tt{\Delta}(0)}=0$) to any output-side semantic basis,
(where $\vec{s}_{\tt{\Delta}(m)}=\vec{s}_{o\tt{(target)}} - \vec{s}_{o\tt{(argmax)}}$).
%
The formula for calculating semantic differences is as follows:

\begin{equation}
\label{eq:linear_interpolation2}
\begin{aligned}
~\vec{s}_{\tt{\Delta}(k)} &= \vec{s}_{k\tt{(target)}} - \vec{s}_{k\tt{(argmax)}}\\
&= \mathtt{scale}_{k}\left(\vec{s}_{o\tt{(target)}} - \vec{s}_{o\tt{(argmax)}}\right)\\
\end{aligned}
\end{equation}



\paragraph{Steering the latent representation between semantic bases}
Based on the semantic difference of two semantic bases in each model layer, the objective of LM repair is to steer the latent representation of the medium token towards the ground truth.
In \cref{fig:perspective}, as shown in the 3rd-layer latent space, a green solid line steers the transition trace from the dark solid curve to the dark dashed curve, which is closer to the green transition route. Further, the latent representation in the last-layer latent space is also affected, moving closer to the ground truth (and the argmax token becomes blue in the vocabulary). The steering change made to the normal transition trace is referred to as ``semantic steer''.
In the process, the steering value used to patch buggy neurons is the normalized value of the semantic difference, denoted as $\vec{s}_{\tt{\Delta}}$. That is, $\vec{s}_{\tt{\Delta}} = \mathtt{norm}(\vec{s}_{\tt{\Delta}(k)})$. Further, the effect of the scaling function is eliminated, as follows:

\begin{equation}
\label{eq:linear_interpolation}
\begin{aligned}
~\vec{s}_{\tt{\Delta}} &= \mathtt{norm}\left(\mathtt{scale}_{k}\left(\vec{s}_{o\tt{(target)}} - \vec{s}_{o\tt{(argmax)}}\right)\right)\\
&= \mathtt{norm}\left(\vec{s}_{o\tt{\left(target\right)}} - ~\vec{s}_{o\tt{\left(argmax\right)}}\right)\\
\end{aligned}
\end{equation}

For each neuron to patch, its new parameters $\vec{r}_{k}{'}$ are the sum of the old parameters $\vec{r}_{k}$ plus the steering value, with a neuron-wise coefficient $\alpha$, which allows each neuron to be updated adaptively.
The coefficient is determined as the minimal value that maximizes the probability of the target token.
The computation formula is as follows:



\begin{equation}
\label{eq:vector_addition}
\begin{aligned}
~\vec{r}_{k}{'} &= \frac{\vec{r}_{k} + \alpha\vec{s}_{\tt{\Delta}}}{1 + \alpha}\\
\end{aligned}
\end{equation}








In the evaluation, we explore two  variants based on the way the neuron patch is estimated:
(1) the first variant only uses the semantic basis of the target token to estimate the neuron patch, rather than the difference of semantic bases, namely $\vec{s}_{\tt{\Delta}}=\mathtt{norm}(\vec{s}_{o\tt{\left(target\right)}})$, labeled as \expttag{[est-basis]};
(2) the second variant eliminates the use of the additional neuron-wise coefficient, namely $\vec{r}_{k}{'}=\vec{r}_{k}+\vec{s}_{\tt{\Delta}}$, labeled \expttag{[est-plain]}.

\subsection{Locating and Patching Buggy Neurons}


The attribution score measures the contribution of neurons in \ul{causing} a model failure, but cannot indicate the benefit of patching each neuron in \ul{solving} the model failure.
We use the term ``patching-gain'' to represent the benefit of patching a specific neuron in repairing model failures.
%
MINT measures the patching-gain by evaluating the impact of making changes to the neuron parameters, known as neuron patches. These patches aim to improve the model's performance by solving failures. The patching-gain will help identify the neurons that are essential in solving model failures, referred to as buggy neurons.
Through the simulation of neuron-patching, we measure the patching-gains of each neuron and plan their priorities to patch.


We simulate the process of neuron-patching progress to locate buggy neurons.
The simulation targets a subset of neurons with the highest attribution scores, as these neurons have the greatest contribution to the failure. However, it's important to note that neurons with high attribution scores aren't always the ones that are effective in solving the failure. This is the reason why it's crucial to further measure the patching-gain of each neuron, to assess the neurons that will be essential in solving the failure once patched.

For each patch applied to a corresponding neuron, we measure its patching-gain by estimating how much it reduces the probability gap between the target token and the argmax token.
The patching-gain measures the impact of making changes to a neuron on improving the probabilities of the target token.
In other words, the patching-gain reflects how effectively the neuron patch enhances the model's ability to rank higher the target token.
The measurement formula of the patching-gain is shown in \cref{eq:measure}, where $p$ indicates the token probability.
Based on the patching-gain, neurons that demonstrate a higher gain from patching will be given higher priority for patching.

\begin{equation}
\label{eq:measure}
\begin{aligned}
\tt{gain}&=p_{\tt{argmax}} - p_{\tt{target}}\\
\end{aligned}
\end{equation}

In the evaluation, we also explore the variant that uses the attribution score as the patching-gain, namely $\tt{gain}=\tt{score}$, which is labeled \expttag{[gain-score]}.

\section{Experiments}
\label{sec:experiments}


\subsection{Research Questions}

To evaluate the performance of MINT, we design a set of experiments to answer the following research questions. 

\begin{reqs}

\item [\req{1}] \textbf{How effective is MINT in repairing Code LMs?}

\end{reqs}



\paragraph{Setup}
We run experiments on the code generation datasets (ref. \cref{subsec:dataset}), and compare the results of our approach with baselines (ref. \cref{subsec:baseline}) before and after model repair.

\paragraph{Measures}
We employ multiple metrics to measure the effectiveness of model repair, and they have different focuses:
\textit{Exact Match} is the proportion of generated tokens that are exactly the same as the ground truth;
\textit{Edit Similarity} is the similarity between the generated tokens and the ground truth based on the minimal number of character transformations; 
\textit{BLEU}~\cite{Papineni2002BleuAM} score is the average percentage of overlapped $n$-gram, typically \num{4}-gram, between the prediction and its ground truth;
while \textit{ROUGE}~\cite{Lin2004ROUGEAP} score is the average percentage of overlapped longest common sequences between the prediction sentence and its ground truth.





\begin{reqs}

\item [\req{2}] \textbf{How efficient is MINT in repairing Code LMs?}

\end{reqs}


\paragraph{Setup}
The main settings are the same as in RQ1. Our approach is compared with the baseline approach \textsc{KN} as described in \cref{subsec:baseline}, with diverse models and datasets.

\paragraph{Measures}
The execution efficiency of model repair is estimated with two metrics.
We count the updated number of neurons that are patched per solved LM failure, denoted as \textit{Updated-Neuron Cost}. The amount of patched neurons indicates the cost of operations, such as locating neurons and generating patches.
Also, we measure the elapsed execution time in seconds per solved LM failure, denoted as \textit{Elapsed-Time Cost}.
For both \textit{updated-neuron cost} and \textit{elapsed-time cost}, smaller values indicate better model-repair efficiency.




\begin{reqs}

\item [\req{3}] \textbf{What is the generalization and specificity of MINT in repairing Code LMs?}

\end{reqs}

\paragraph{\textbf{Generalization \& Specificity}}
In the context of LM repair, generalization refers to the ability of model repair methods to identify the related data (which shares the same ground truth) and maximize the impact there;
Specificity refers to the ability of model repair methods to identify the unrelated data (with different ground truth) and minimize the impact there.
% ...
There is a balance between generalization and specificity, over-fitting suggests the case of good generalization and bad specificity, and under-fitting suggests the case of bad generalization and good specificity.



\paragraph{Setup}
The experiments are conducted on a new benchmark (ref. \cref{subsec:benchmark}) to compare our approach with its variants and the baselines, concerning generalization and specificity.




\paragraph{Measures}
Given that the goal of model repair is to elevate the target token to the argmax token, we measure the change in the probability gap between the argmax token and the target token.
To measure the impact of model repair on the probability gap, we measure the changes to the accuracy, noted as $\Delta\text{Acc}$.
In addition, we compute \textit{Mean Absolute Error (MAE)} and \textit{Root Mean Square Error (RMSE)}.
The scores on related data (those sharing the same target token) are generalization scores, and the scores on unrelated data (with different target tokens) are specificity scores.




In terms of generalization, the positive effects on related data should be maximal. This leads to substantial changes in the probability gap (between the argmax token and the target token). Strong generalization is indicated by significant improvements in accuracy, along with high MAE and RMSE scores.
In contrast, when evaluating specificity, the negative effects on unrelated data should be minimal. This means that the changes in the probability gap should be small. Good specificity is characterized by minimal changes in accuracy (indicating stable accuracy), and low MAE and RMSE scores.






\subsection{Design of Experiments}
\label{sec:exp:overview}

\begin{figure}[htbp]
    \centering
    \includegraphics[width=1.0\linewidth]{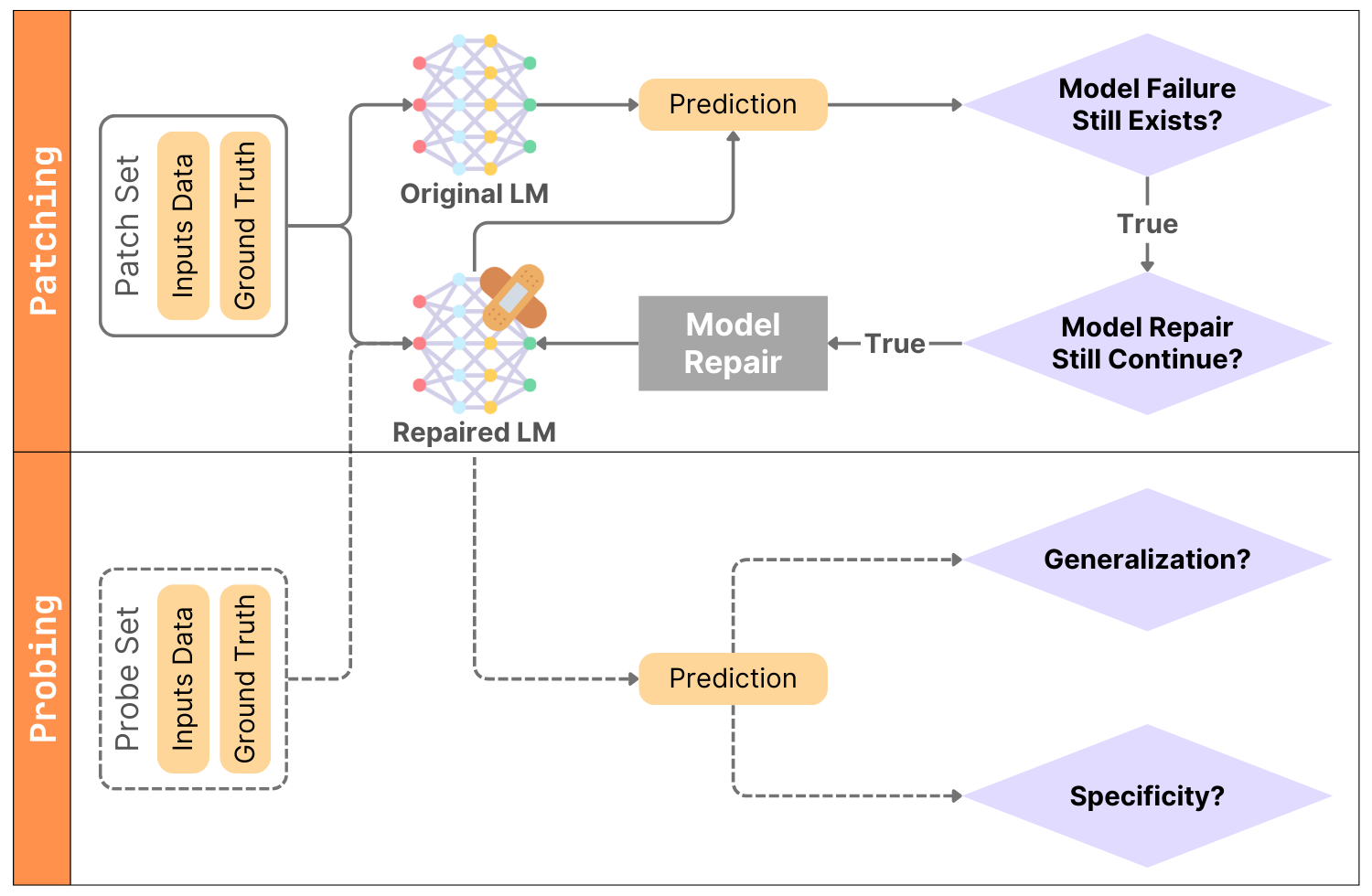}
    \caption{Overview of experimental design for model repair.}
    \label{fig:overview}
\end{figure}

As the overview of experiments shown in \cref{fig:overview}, we designed two types of experiments: ``patching'' and ``probing''.
The former is for RQ1 and RQ2, to study the effectiveness and efficiency of LM repair, while the latter is for RQ3, to study the generalization and specificity.

In ``patching'' experiments, let's assume an LM fails to predict the next token correctly, that is, the target token and the argmax token are not the same. We perform model repair by patching buggy neurons to get the repaired model, and then, we employ model predictions to validate whether the instance of the LM failure is solved.
The procedure of patching and validating with the same data is similar to program repair. When a test case exposes a program failure, the failure is identified, and subsequent patching is done to solve it. The expectation is that the patched program should then pass the same test case after the repair. We call each input-output pair a ``patch set''.
The code generation datasets used for the experiments are described in~\cref{subsec:dataset}.

In ``probing'' experiments, we assess how model repair affects other instances of the same LM failure, while ensuring that the repair is focused on the specific instance at hand.
The ideal outcome is achieved when other instances of the failure are corrected, and concurrently, other knowledge and behaviors of the LM remain unaffected.
This balance between generalization and specificity is key to ensuring the repaired LM performs correctly across different scenarios.
A set of input-output pairs is referred to as a ``probe set''. To conduct these experiments, we design a dedicated benchmark that incorporates related data for generalization and unrelated data for specificity, described in \cref{subsec:benchmark}.


\paragraph{\textbf{Patch Set \& Probe Set}}
Different from the concept of \textit{training set} used in model training, only a few exemplary data are available in \textit{patch set}, and they are used to identify and solve model failures.
Similar to the concept of \textit{test set} in model evaluation, the additional data from \textit{probe set} are used to evaluate generalization or specificity, by probing model knowledge and behaviors.
Corresponding to one patch set, there are two probe sets (one is to study generalization, and one is to study specificity).
We make sure that there is \emph{no overlap} between the patch set and the two probe sets.
The concepts of training set and test set are employed in model training and testing, while patch set and probe set are in the context of model repair, hence we use different terms.

\subsection{Datasets}
\label{subsec:dataset}




We employ the following code generation datasets to answer RQ1 and RQ2. They cover different popular coding tasks (Python, assembly, bash code), with diverse stats (data length, task difficulty, etc).
The dataset statistics are shown in \cref{tab:corpus}.

\begin{enumerate}
    \item \textit{Line-level Code Generation:} We employ \textbf{CoNaLa}~\cite{Yin2018LearningTM}, which consists of pairs of rewritten intents (interline comments) and lines of Python code. Given a natural language intent, the model generates the most suitable code. The data is extracted from top-scored answers in top-viewed Stack Overflow posts, dated March 2017;
    \item \textit{Shellcode Generation:} We employ \textbf{IA32}~\cite{Liguori2021CanWG} which consists of pairs of natural language intents and assembly code (for Intel Architecture, 32-bit). Given a natural language intent, the model generates the correcting assembly code. The data collects 20 years of shellcodes from a variety of sources, annotated with descriptions of a typical style;
    \item \textit{Intent-to-Bash Translation:} We employ \textbf{TLDR}~\cite{Zhou2022DocCoderGC} dataset which is composed of pairs of natural language intents and bash commands. Given an intent, the model generates the corresponding bash command. The data is harvested from the English subset of the tldr-pages project~\footnote{https://tldr.sh/}, collaborative cheatsheets for console commands.
\end{enumerate}

\begin{table}[!ht]
    \caption{Statistics of Code Generation Datasets.}
    \label{tab:corpus}
    \centering
    \resizebox{0.8\linewidth}{!}{
        \sisetup{table-format=6}
\begin{tabular}{
    ll rrr
}

\toprule

& & \textbf{CoNaLa} & \textbf{IA32} & \textbf{TLDR} \\





\midrule
\multirow{2}*{Data Num.}
& Retrieval & 2,379 & 2,560 & 6,414 \\
& Inference & 500 & 640 & 928 \\

\midrule
\multirow{2}*{Avg. Length}
& Inputs & 59.9 & 47.7 & 48.4 \\
& Outputs & 40.4 & 16.6 & 35.3 \\

\bottomrule

\end{tabular}

 }
\end{table}

\paragraph{Usage}
We iterate the data in the inference division and treat each of them as a patch set (the size is always $1$).
If a patch set contains LM failure(s), we do model repair with its elements to solve it, and validate that the failure is indeed solved with the same elements.
The retrieval division is the corpus to retrieve demonstrations for in-context prompting~\cite{Dong2022ASF}.


\subsection{Benchmark}
\label{subsec:benchmark}

We develop a new benchmark for evaluating the generalization and specificity of patching code LMs in \req{3}.
From the benchmark, the related data and unrelated data are adequate in quantity, extensive in characteristics, and can be collected flexibly.
In contrast, existing datasets cannot satisfy the requirements to study the side effects of LM repair.

The benchmark has a total size of $450$ data samples. As shown in \cref{tab:benchmark}, it covers $3$ types and $15$ subtypes. In each subtype, there are $30$ data samples, that is $10$ crafted samples times $3$ different variants.
The dataset covers a wide range of LM failures with diverse properties, including operators, keywords, and API names of different functionalities.
In each subtype of the benchmark, we prepare $3$ pairs of argmax tokens and target tokens (one fixed argmax token, pairing with three different target tokens).
For each pair of tokens, we manually craft $10$ different data samples that contain the argmax tokens, covering the common usages of corresponding properties.
In addition, there are demonstrative input-output pairs in the benchmark for in-context promptings.


In next-token prediction tasks, a particular model failure is solved in a manner that promotes the target token to its argmax token.
With reference to the data used for model repair, we collect the related data (which shares the same ground truth) from the benchmark to assess generalization, but collect the unrelated data (with different ground truth) to assess specificity.
For generalization, we anticipate the related data will also exhibit the target token being closer to or even identical to the argmax token;
While for specificity, we don't anticipate the unrelated data to have the target token closer to or matching the argmax token.

\paragraph{Usage}
We iterate the benchmark and let each datum be the \textit{patch set} to do model repair, and dynamically collect related data (when studying generation) or unrelated data (when studying specificity) from the benchmark as the \textit{probe sets}.


In the benchmark, for a datum to undergo model repair, we collect the data from the same subtype as the probe set for generalization study, noted as $G$. Since they have the same argmax token, they tend to share the same knowledge, they will need the same target token to be the new argmax token.
In contrast, we collect the data from different subtypes as the probe set for specificity study, noted as $S$. The reason is that, the data having no common argmax token are hard to express the same knowledge, so they won't expect the current target token to be the new argmax token.

\begin{table}[!tb]
    \caption{Summary of the Benchmark Dataset.}
    \label{tab:benchmark}
    \centering
    \resizebox{1.0\linewidth}{!}{
        \sisetup{table-format=2.2}
\begin{tabular}{
    ll ll r
}

\toprule

\textbf{Type} & \textbf{SubType} & \textbf{ArgmaxToken} & \textbf{TargetToken} & \textbf{\#} \\

\midrule

\multirow{5}{*}{\rotatebox[origin=c]{90}{\expttag{Expression}}}
& {assign} & {\verb|=|} & {\verb|+|, \verb|~|, \verb|%|} & 10 \\
& {compound} & {\verb|*=|} & {\verb|+=|, \verb|/=|, \verb|!=|} & 10 \\
& {arithmetic} & {\verb|%|} & {\verb|+|, \verb|*|, \verb|/|} & 10 \\
& {bit\_logical} & {\verb|&|} & {\verb|<<|, \verb|~|, \verb|^|} & 10 \\
& {comparison} & {\verb|>=|} & {\verb|<=|, \verb|==|, \verb|!=|} & 10 \\

\midrule

\multirow{5}{*}{\rotatebox[origin=c]{90}{\expttag{Statement}}}
& {jump\_ret} & {`return'} & {`break', `pass', `yield'} & 10 \\
& {loop\_for} & {`for'} & {`if', `match', `while'} & 10 \\
& {cond\_if} & {`if'} & {`case', `elif', `for'} & 10 \\
& {cond\_and} & {`and'} & {`not', `is', `or'} & 10 \\
& {define\_def} & {`def'} & {`class', `lambda', `partial'} & 10 \\

\midrule

\multirow{5}{*}{\rotatebox[origin=c]{90}{\expttag{Invocation}}}
& {std\_abs} & {`abs'} & {`hex', `max', `round'} & 10 \\
& {std\_type} & {`type'} & {`bool', `hash', `len'} & 10 \\
& {string} & {`count'} & {`find', `replace', `split'} & 10 \\
& {numpy} & {`choice'} & {`normal', `sample', `shuffle'} & 10 \\
& {pandas} & {`to\_json'} & {`to\_csv', `to\_pickle', `to\_sql'} & 10 \\

\bottomrule

\end{tabular}

 }
\end{table}

\paragraph{Example}
Let's assume a datum from the ``assign'' subtype is the patch set for model repair, we take the other data in the same subtype as the probe set $G$ (for generalization), containing $9$ elements; and collect the data from all other subtypes as the probe set $S$ (for specificity), containing $14$ elements.
When the target token `\verb|+|' is the argmax token in model repair, we expect the same effects on other data of the ``assign'' subtype, but we won't expect it to affect the data of other subtypes.
For the same subtype, the performance in other cases will also be studied, where the target token is `\verb|~|' or `\verb|%|'.
Similarly, the computation will repeat on all $450$ data samples.
The scores will be averaged as an overall measure of the generalization and the specificity.


\subsection{Models}
\label{subsec:model}

The models used in our experiments are \textsc{CodeGen-2B}~\cite{Nijkamp2022CodeGenAO}, \textsc{StarCoder2-3B}~\cite{Lozhkov2024StarCoder2A}, and \textsc{CodeLlama-7B}~\cite{Rozire2023CodeLO}.
They are competitive and modern open-source language models for code generation, and with diverse stats (release date, data scale, etc).
They are selected based on the popularity (especially, most downloads per month) in the hugging-face website\footnote{https://huggingface.co/}. As time goes on, the popularity gradually varies.
The model details are shown in \cref{tab:model}.
\begin{table}[!ht]
    \caption{Statistics of Code Language Models.}
    \label{tab:model}
    \centering
    \resizebox{0.85\linewidth}{!}{
        \sisetup{table-format=6}
\begin{tabular}{
    l rrr
}

\toprule

& \textbf{CodeGEN-2B} & \textbf{StarCoder2-3B} & \textbf{CodeLlama-7B} \\

\midrule

Release Date & Oct, 2022 & Feb, 2024 & Aug. 2023 \\

\midrule

Data Scale & 0.5T tokens & 3T tokens & 0.5T tokens \\



\midrule

Layer Num. & 32 & 30 & 32 \\

\midrule

Hidden Size & 2,560 & 3,072 & 4,096 \\



\midrule

Vocabulary & 51,200 & 49,152 & 32,016 \\

\bottomrule

\end{tabular}

 }
\end{table}

\subsection{Baselines}
\label{subsec:baseline}

To the best of our knowledge, the \ac{sota} in model repair for next-token prediction is \textsc{KN}~\cite{Dai2021KnowledgeNI}.
\textsc{KN} takes a ``locate-then-update'' practice:
it locates the critical neurons based on the attribution analysis on parallel data, since the parallel data tend to share the same critical neurons.
Once located the neurons, \textsc{KN} simply strengthens and weakens their activations, instead of updating neuron parameters.



\textsc{KN} was proposed for \textsc{BERT} models so cannot be directly used in generative tasks, we adapted \textsc{KN} to support generative models.
Besides, different from \textsc{MINT}, \textsc{KN} requires additional parallel data in locating critical neurons.
We retrieve semantically similar data from the corpus, and their ground truth is the same as the current datum.


\subsection{Implementation Details}
\label{sec:exp:details}

For in-context prompting, we retrieve the most semantically similar input-output pair and take it as the one-shot demonstration.
In the model repair, the number quota of neurons to patch in solving a model failure is $5$.
For our approach, the number of neurons chosen to compute their patching-gains is $10$.
For the baseline \textsc{KN}, the amount of the required parallel data for locating buggy neurons is $10$, as recommended.
Experiments were conducted on one Nvidia A100 GPU.


\section{Results}
\label{sec:results}

In the study of \req{1} and \req{2}, optimal scores are highlighted in bold.
In the \req{3} study, the results of the optimal balance between generalization and specificity are highlighted in bold.
To facilitate the analysis, the scores of baselines and variants better than \textsc{MINT} are emphasized in grey color.

\subsection{\req{1} Results}
\label{subsec:rq1_results}

\begin{table*}[!ht]
    \caption{Comparison of Effectiveness of Model Repair for Code Generation.}
    \label{tab:results_rq1_effects}
    \centering
    \resizebox{0.95\linewidth}{!}{%
        \sisetup{table-format=2.2}
\begin{tabular}{
    l rrr rrr rrr rrr
}

\toprule

\multirow{2}[2]{*}{\textbf{Approach}} & \multicolumn{3}{c}{\textbf{ExactMatch $\uparrow$}} & \multicolumn{3}{c}{\textbf{EditSimi $\uparrow$}} & \multicolumn{3}{c}{\textbf{BLEU $\uparrow$}} & \multicolumn{3}{c}{\textbf{ROUGE $\uparrow$}} \\
\cmidrule(lr){2-4} \cmidrule(lr){5-7} \cmidrule(lr){8-10} \cmidrule(lr){11-13}
& {\textbf{CoNaLa}} & {\textbf{IA32}} & {\textbf{TLDR}}
& {\textbf{CoNaLa}} & {\textbf{IA32}} & {\textbf{TLDR}}
& {\textbf{CoNaLa}} & {\textbf{IA32}} & {\textbf{TLDR}}
& {\textbf{CoNaLa}} & {\textbf{IA32}} & {\textbf{TLDR}} \\


\midrule
\expttag{\textsc{CodeGen-2B}}
& 0.762 & 0.913 & 0.513
& 0.820 & 0.934 & 0.608
& 0.655 & 0.889 & 0.399
& 0.766 & 0.914 & 0.496 \\
\expttag{+~\textsc{KN}}
& 0.779 & 0.918 & 0.538
& 0.833 & 0.937 & 0.622
& 0.675 & 0.898 & 0.423
& 0.779 & 0.920 & 0.507 \\
\expttag{+~\textsc{MINT}}
& \textbf{0.870} & \textbf{0.974} & \textbf{0.689}
& \textbf{0.901} & \textbf{0.980} & \textbf{0.742}
& \textbf{0.769} & \textbf{0.960} & \textbf{0.476}
& \textbf{0.873} & \textbf{0.976} & \textbf{0.693} \\

\midrule
\expttag{\textsc{StarCoder2-3B}}
& 0.810 & 0.922 & 0.619
& 0.850 & 0.943 & 0.690
& 0.727 & 0.916 & 0.502
& 0.806 & 0.919 & 0.599 \\
\expttag{+~\textsc{KN}}
& 0.828 & 0.926 & 0.653
& 0.864 & 0.945 & 0.713
& 0.752 & 0.925 & 0.549
& 0.823 & 0.924 & 0.621 \\
\expttag{+~\textsc{MINT}}
& \textbf{0.935} & \textbf{0.983} & \textbf{0.861}
& \textbf{0.949} & \textbf{0.986} & \textbf{0.885}
& \textbf{0.889} & \textbf{0.979} & \textbf{0.727}
& \textbf{0.937} & \textbf{0.985} & \textbf{0.868} \\

\midrule
\expttag{\textsc{CodeLlama-7B}}
& 0.773 & 0.825 & 0.580
& 0.824 & 0.881 & 0.651
& 0.666 & 0.696 & 0.469
& 0.767 & 0.831 & 0.555 \\
\expttag{+~\textsc{KN}}
& 0.797 & 0.838 & 0.624
& 0.842 & 0.890 & 0.685
& 0.696 & 0.706 & 0.511
& 0.793 & 0.842 & 0.590 \\
\expttag{+~\textsc{MINT}}
& \textbf{0.885} & \textbf{0.877} & \textbf{0.809}
& \textbf{0.909} & \textbf{0.917} & \textbf{0.840}
& \textbf{0.809} & \textbf{0.825} & \textbf{0.663}
& \textbf{0.888} & \textbf{0.901} & \textbf{0.815} \\

\bottomrule

\end{tabular}
 }
\end{table*}


\textsc{MINT} shows significant improvements over the baseline \textsc{KN} with respect to all metrics, in all three code generation tasks and all LMs, as shown in \cref{tab:results_rq1_effects}. Based on our analysis, the main reason is that the baseline \textsc{KN} often fails to apply useful changes to neurons, even though it uses 10 times the amount of data to locate the buggy neurons.
In contrast, the semantic-based patches in our approach \textsc{MINT} are more effective in improving the probabilities of the target token.

The performance of language models is different in code generation: \textsc{StarCoder2-3B} performs the best in all tasks, then \textsc{CodeLlama-7B}, while \textsc{CodeGEN-2B} performs the worst. The release dates of LMs influence their performance, as more recent models are trained on larger and better-quality datasets, which may enhance their capabilities, although model size also plays a significant role. However, in shellcode generation (the IA32 dataset), \textsc{CodeGEN-2B} performs better than \textsc{CodeLlama-7B}. 
The variation in performance could be due to the task complexity. For simpler tasks, smaller models might outperform larger ones because they require fewer parameters and are less prone to overfitting or unnecessary complexity.
The effects of model repair on language models appear to vary. The effects on smaller LMs, such as \textsc{CodeGEN-2B} and \textsc{StarCoder2-3B}, are similar, but not as significant as on \textsc{CodeLlama-7B}. Intuitively, this difference is caused by the amount of parameters, since a larger model with more parameters will have more neurons to patch.


The effectiveness of model repair differs across tasks. For tasks with less difficulty (where the length of ground truth is short), models perform well so the improvements by different approaches are not that huge, but for challenging tasks, the effects of model repair techniques are more obvious. For example, with \textsc{StarCoder2-3B}, BLEU scores in IA32 increase from $0.916$ to $0.979$, but MINT improves the score in CoNaLa from $0.727$ to $0.889$, and the score increases in TLDR from $0.502$ to $0.727$. Furthermore, some metrics are sensitive to the improvements achieved by \textsc{KN} and \textsc{MINT}. For example, with \textsc{StarCoder2-3B}, the ROUGE score in CoNaLa rise by $1.7\%$ for KN, and $13.1\%$ for \textsc{MINT}, in IA32 the ROUGE score increases by $0.5\%$ and $6.6\%$ for \textsc{KN} and \textsc{MINT} respectively, and in TLDR the ROUGE score increases by $2.2\%$ for \textsc{KN} and $26.9\%$ for \textsc{MINT}.
In contrast, EditSimi produces similar scores for both two approaches, indicating that it is less sensitive to the specific improvements introduced by \textsc{KN} and \textsc{MINT}.



\subsection{\req{2} Results}
\label{subsec:rq2_results}

Compared to the \textsc{KN} baseline, \textsc{MINT} is efficient in locating buggy neurons and effective in patching buggy neurons.
As shown in \cref{tab:results_rq2_costs}, in terms of updated-neuron cost, our approach \textsc{MINT} only patches $1$ neuron while \textsc{KN} changes on average $2$ neurons for each solved failure; in terms of elapsed-time cost, our approach tends to require shorter time for processing all LM failures than the baseline.


\begin{table}[!ht]
    \caption{Comparison of Efficiency of Model Repair.}
    \label{tab:results_rq2_costs}
    \centering
    \resizebox{1.0\linewidth}{!}{%
        \sisetup{table-format=2.2}
\begin{tabular}{
    l rrr rrr
}

\toprule

\multirow{2}[2]{*}{\textbf{Approach}}
& \multicolumn{3}{c}{\textbf{\# Updated-Neuron $\downarrow$}}
& \multicolumn{3}{c}{\textbf{\# Elapsed-Time $\downarrow$}} \\
\cmidrule(lr){2-4} \cmidrule(lr){5-7}
& \textbf{CoNaLa} & \textbf{IA32} & \textbf{TLDR}
& \textbf{CoNaLa} & \textbf{IA32} & \textbf{TLDR} \\

\midrule

\expttag{\textsc{CodeGen-2B}}
& -- & -- & --
& -- & -- & -- \\
\expttag{+~\textsc{KN}}
& 2.17 & 1.61 & 2.14
& 31.4 s & 32.7 s & 26.2 s \\
\expttag{+~\textsc{MINT}}
& \textbf{1.00} & \textbf{1.00} & \textbf{1.00}
& \textbf{15.8 s} & \textbf{22.0 s} & \textbf{14.0 s} \\

\midrule

\expttag{\textsc{StarCoder2-3B}}
& -- & -- & --
& -- & -- & -- \\
\expttag{+~\textsc{KN}}
& 2.50 & 1.90 & 2.28
& 30.6 s & 42.4 s & 23.4 s \\
\expttag{+~\textsc{MINT}}
& \textbf{1.00} & \textbf{1.00} & \textbf{1.00}
& \textbf{18.9 s} & \textbf{31.7 s} & \textbf{15.7 s} \\

\midrule

\expttag{\textsc{CodeLlama-7B}}
& -- & -- & --
& -- & -- & -- \\
\expttag{+~\textsc{KN}}
& 2.27 & 2.27 & 2.19
& 38.1 s & 40.0 s & 33.9 s \\
\expttag{+~\textsc{MINT}}
& \textbf{1.00} & \textbf{1.00} & \textbf{1.00}
& \textbf{12.5 s} & \textbf{12.9 s} & \textbf{11.8 s} \\

\bottomrule

\end{tabular}
 }
\end{table}

In addition, the baseline \textsc{KN} is more likely to skip repairing models compared to our approach \textsc{MINT}.
Frequent skips may be caused by two reasons: the locating method is ineffective, or the updating method is ineffective.
Our observations indicate that \textsc{KN} has limitations in its approach to neuron patching, as it relies on a simplistic mechanism of merely doubling the activations of the identified neurons. This method lacks robust theoretical backing or empirical evidence to demonstrate its effectiveness in improving model performance or repairing incorrect model behavior.
In contrast, our approach is theoretically supported by vocabulary-defined semantics~\cite{Gu2024VocabularyDefinedSL}.
Besides, unlike \textsc{KN}, \textsc{MINT} typically requires fewer neurons to correct the next-token prediction. This difference arises from the limited neuron-patching capability of \textsc{KN}.
Also, the baseline \textsc{KN} does not optimize the order in which neurons are patched, which may be one more reason for its higher neuron cost.

Results also show that \textsc{KN} requires more time to process each LM failure, which is mainly due to the ineffective attribution method.
The time cost of model repair is generally for two purposes: locating neurons and patching neurons.

In our experiments, the reason why \textsc{KN} takes a longer time is because it locates neurons several times with the required parallel data.
The attribution method used \textsc{MINT} is slightly slower than that used in \textsc{KN}, so when locating for the same amount of times, our approach is slower. However, our approach locates buggy neurons when knowing how to patch neurons, so is more targeted.
Meanwhile, the average time cost for patching each neuron can be very close since the practices of \textsc{MINT} and \textsc{KN} are both direct. The former is modifying the neuron parameters while the latter is modifying the neuron activations.

Furthermore, \textsc{MINT} demonstrates low variability in elapsed-time costs, highlighting its consistent and stable capability for solving model failures. For instance, when comparing \textsc{CodeGEN-2B} and \textsc{StarCoder2-3B}, the \textsc{KN} method exhibits mixed performance: it completes repairs more quickly on the IA32 dataset but takes longer on other datasets. In contrast, \textsc{MINT} consistently outperforms \textsc{KN} by taking less time across all datasets. This consistency arises because \textsc{MINT} rarely skips repairing models, resulting in a roughly uniform time cost for addressing each LM failure.




\subsection{\req{3} Results}
\label{subsec:rq3_results}


In \req{3}, we evaluate the generalization and specificity of the model repair approaches using two metrics: $\Delta\text{Acc}$ and MAE/RMSE scores.
Better generalization corresponds to higher scores, while better specificity corresponds to lower scores.
$\Delta\text{Acc}$ quantifies the direct changes in accuracy for token predictions. It highlights how the corrections have impacted the model's outputs.
MAE/RMSE metrics, on the other hand, measure the overall statistical changes in token probabilities. They reflect the magnitude of adjustments in token probabilities.
Together, these metrics provide complementary insights. $\Delta\text{Acc}$ focuses on the tangible improvements in prediction accuracy, while MAE/RMSE reveals broader trends in the model's behavior.
In the context of model repair, generalization and specificity are orthogonal, hard to be optimal simultaneously, and a trade-off between them must be considered.
The balance between the generalization and specificity scores is evaluated using an empirical formula: $\tt{ratio}=\tt{generalization}/\tt{specificity}$, which means the growth in generalization achieved while tolerating the loss of unit specificity.
A larger ratio indicates a better balance, while the under-fitting cases won't be considered due to the numerical distortion caused by division-by-zero.
Based on the scores in \cref{tab:results_rq3_reliability}, our approach \textsc{MINT} is generally the most balanced, with high generalization scores and low specificity scores. However, in \textsc{StarCoder2-3B}, \textsc{MINT} is suboptimal, since its ratios corresponding to $\Delta\text{Acc}$ and MAE/RMSE scores are $8.207$, $7.373$, $3.335$ while the scores of the variant \expttag{[gain-score]} are $9.381$, $8.361$, $3.653$.

Based on the scores in \cref{tab:results_rq3_reliability}, we can study the extreme cases of over-fitting and under-fitting.
The variant \expttag{[est-plain]} indicates a severe over-fitting case. It has high generalization scores and high specificity scores. The key characteristic of \expttag{[est-plain]} is that it omits the use of a coefficient during neuron-patching. As a result, the updates to neuron parameters are more substantial compared to \textsc{MINT}. The larger updates make it challenging to appropriately solve the model failures. Additionally, the excessive parameter changes impair the overall capability of the LM.
The baseline \textsc{KN} and the variant \expttag{[attr-rand]} are prone to under-fitting, as both exhibit low generalization and low specificity scores. Despite these similarities, their performance differs significantly in addressing model failures: \textsc{KN} frequently fails to solve model failures while \expttag{[attr-rand]} consistently solves them.
It indicates that the baseline is not that effective and reliable in repairing models. Also, the changes to neurons tend to compensate for each other, and patching non-critical neurons may cause side effects.

Unlike other variants, \expttag{[est-basis]} stands out due to its unique performance. It has lower generalization scores and higher specificity scores compared to \textsc{MINT}, indicating an overall weaker capability in both dimensions.
\expttag{[est-basis]} uses the semantic basis of the target token to estimate neuron patches, instead of the semantic differences between the target token and the argmax token.
The results indicate that, the semantics of both tokens are essential to model repair.


\begin{table}[!ht]
    \caption{Comparison of the Generalization and Specificity.}
    \label{tab:results_rq3_reliability}
    \centering
    \resizebox{1.0\linewidth}{!}{%
        \sisetup{table-format=2.2}
\begin{tabular}{
    l rrr rrr
}

\toprule

\multirow{2}[2]{*}{\textbf{Approach}} & \multicolumn{3}{c}{\textbf{Generalization $\uparrow$}} & \multicolumn{3}{c}{\textbf{Specificity $\downarrow$}} \\
\cmidrule(lr){2-4} \cmidrule(lr){5-7}
& \textbf{$\Delta\text{Acc}$} & \textbf{$\text{MAE}$} & \textbf{$\text{RMSE}$}
& \textbf{$\Delta\text{Acc}$} & \textbf{$\text{MAE}$} & \textbf{$\text{RMSE}$} \\


\midrule

\expttag{\textsc{CodeGen-2B}}
& -- & -- & --
& -- & -- & -- \\
\expttag{+~\textsc{KN}}
& 0.000 & 0.000 & 0.000
& \tabh 0.000 & \tabh 0.000 & \tabh 0.000 \\
\expttag{+~\textsc{MINT}}
& \textbf{0.675} & \textbf{0.295} & \textbf{0.436}
& \textbf{0.156} & \textbf{0.040} & \textbf{0.113} \\
\expttag{+~[attr-actv]}
& 0.556 & \tabh 0.329 & \tabh 0.479
& 0.183 & 0.114 & 0.246 \\
\expttag{+~[attr-rand]}
& 0.008 & 0.019 & 0.062
& \tabh 0.001 & \tabh 0.004 & \tabh 0.025 \\
\expttag{+~[est-basis]}
& 0.533 & 0.205 & 0.325
& 0.200 & 0.052 & 0.151 \\
\expttag{+~[est-plain]}
& \tabh 0.892 & \tabh 0.333 & \tabh 0.476
& 0.353 & 0.100 & 0.234 \\
\expttag{+~[gain-score]}
& 0.664 & \tabh 0.297 & \tabh 0.440
& 0.158 & 0.043 & 0.125 \\

\midrule

\expttag{\textsc{StarCoder2-3B}}
& -- & -- & --
& -- & -- & -- \\
\expttag{+~\textsc{KN}}
& 0.000 & 0.000 & 0.000
& \tabh 0.000 & \tabh 0.000 & \tabh 0.000 \\
\expttag{+~\textsc{MINT}}
& 0.476 & 0.494 & 0.597
& 0.058 & 0.067 & 0.179 \\
\expttag{+~[attr-actv]}
& \tabh 0.616 & \tabh 0.586 & \tabh 0.669
& 0.121 & 0.130 & 0.267 \\
\expttag{+~[attr-rand]}
& 0.001 & 0.018 & 0.054
& \tabh 0.000 & \tabh 0.002 & \tabh 0.009 \\
\expttag{+~[est-basis]}
& 0.407 & 0.357 & 0.479
& 0.126 & 0.101 & 0.236 \\
\expttag{+~[est-plain]}
& \tabh 0.762 & \tabh 0.576 & \tabh 0.665
& 0.180 & 0.128 & 0.272 \\
\expttag{+~[gain-score]}
& \textbf{0.197} & \textbf{0.301} & \textbf{0.431}
& \tabh \textbf{0.021} & \tabh \textbf{0.036} & \tabh \textbf{0.118} \\

\midrule

\expttag{\textsc{CodeLlama-7B}}
& -- & -- & --
& -- & -- & -- \\
\expttag{+~\textsc{KN}}
& 0.000 & 0.000 & 0.000
& \tabh 0.000 & \tabh 0.000 & \tabh 0.000 \\
\expttag{+~\textsc{MINT}}
& \textbf{0.273} & \textbf{0.300} & \textbf{0.395}
& \textbf{0.053} & \textbf{0.044} & \textbf{0.134} \\
\expttag{+~[attr-actv]}
& \tabh 0.306 & \tabh 0.361 & \tabh 0.441
& 0.078 & 0.065 & 0.159 \\
\expttag{+~[attr-rand]}
& 0.000 & 0.008 & 0.019
& \tabh 0.001 & \tabh 0.001 & \tabh 0.011 \\
\expttag{+~[est-basis]}
& 0.203 & 0.249 & 0.348
& 0.062 & 0.059 & 0.156 \\
\expttag{+~[est-plain]}
& \tabh 0.438 & 0.284 & \tabh 0.415
& 0.096 & 0.064 & 0.173 \\
\expttag{+~[gain-score]}
& 0.040 & \tabh 0.305 & 0.395
& \tabh 0.000 & 0.054 & 0.170 \\

\bottomrule

\end{tabular}

 }
\end{table}

\paragraph{\textbf{Generalization}}
For generalization, higher scores indicate stronger positive effects on related data, as shown in \cref{tab:results_rq3_reliability}.
In all three models, similar to \textsc{MINT}, $2$ out of $5$ variants can maximize their effects in improving the accuracy, and $3$ variants can maximize their effects in the probabilities of the target token.
The overlapped variants, \expttag{[attr-actv]} and \expttag{[est-plain]}, show competitive generalization.

Compared with \textsc{MINT}, the variant \expttag{[attr-actv]} merely uses the activation signals to locate buggy neurons, instead of considering both gradients and activations. Usually, gradients represent the feedback from the outputs while activations capture the input features.
We deduce that the neurons serving common knowledge (such as the code syntax) tend to be located, since they are likely to cause larger activation signals.
As discussed, \expttag{[est-plain]} is an over-fitting case, showing better generalization but worse specificity.

\paragraph{\textbf{Specificity}}
For specificity, higher scores indicate stronger negative effects on unrelated data, as shown in \cref{tab:results_rq3_reliability}.
In all three models, besides the baseline \textsc{KN}, $2$ out of $5$ variants have lower scores than \textsc{MINT}. They are \expttag{[attr-rand]} and \expttag{[gain-score]}, and they show competitive specificity.

Compared with \textsc{MINT}, the variant \expttag{[gain-score]} directly uses the attribution scores to estimate the patching-gain of buggy neurons.
Since attribution scores are not precise in revealing the patching-gain,
this potentially causes more neurons to be patched, leading to low scores in both generalization and specificity. It is similar to the variant \expttag{[attr-rand]}, even though not that exaggerated.
As discussed, \expttag{[attr-rand]} is an under-fitting case, showing worse generalization but better specificity.

\section{Discussion}
\label{sec:discussion}

We present insights on the potential side effects of language model repair with actual cases from our experiments, including the capabilities of LM repair when dealing with multiple repairs; and the efforts of guaranteeing the balance between the generalization and specificity.
We aim for this discussion to inspire future research directions. 

\subsection{Multiple Repairs in Succession}

\textsc{MINT} operates in a mode where it performs successive LM repairs, as outlined in \cref{algo:repair}. Based on the results in \cref{subsec:rq1_results}, it is capable of solving multiple model failures in succession without causing significant side effects.
The potential side effects of multiple repairs may include impacts on the effectiveness of LM repair methods or on the model's performance in code generation (or next-token prediction).
To demonstrate the common limitations of LM repair, we select a few cases from the experiments with \textsc{StarCoder2-3B}, as shown in \cref{tab:showcase}.
\expttag{Inputs} means the natural language prompt, and \expttag{Outputs} means the corresponding ground truth.
\exptlabel{Gen}{before} and \exptlabel{Gen}{after} are the predicted token sequence before and after model repair, while \exptlabel{Gen}{during} indicates the predicted token sequence during model repair (where model repair is conducted for each incorrect token-prediction).
By comparing with the ground truth, the incorrect tokens are highlighted in red while other predicted tokens are in blue.

\begin{table}[!ht]
    \caption{Showcase of Multiple Repairs in Succession.}
    \label{tab:showcase}
    \centering
    \resizebox{\linewidth}{!}{%
        \sisetup{table-format=2.2}
\rowcolors{2}{}{gray!10}
\begin{tabular}{
    cll
}

\hiderowcolors
\toprule

& \textbf{Data} & \textbf{String and Token Sequence} \\

\midrule


\multirow{5}{*}{\rotatebox[origin=c]{90}{\expttag{CoNaLa\#016}}}
& \expttag{Inputs} & {get the first object from a queryset in django model `Entry`} \\
& \expttag{Outputs} & {Entry.objects.filter()[:1].get()} \\
& \exptlabel{Gen}{before} & \hlcyan{Entry}~\hlcyan{.}~\hlcyan{objects}~\hlcyan{.}~\hlpink{all}~\hlpink{(}~\hlcyan{[:}~\hlcyan{1}~\hlpink{]}~\hlcyan{get}~\hlcyan{()} \\
& \exptlabel{Gen}{during} & \hlcyan{Entry}~\hlcyan{.}~\hlcyan{objects}~\hlcyan{.}~\hlcyan{filter}~\hlpink{().}~\hlcyan{[:}~\hlcyan{1}~\hlcyan{].}~\hlcyan{get}~\hlcyan{()} \\
& \exptlabel{Gen}{after} & \hlcyan{Entry}~\hlcyan{.}~\hlcyan{objects}~\hlcyan{.}~\hlcyan{filter}~\hlpink{().}~\hlcyan{[:}~\hlcyan{1}~\hlcyan{].}~\hlcyan{get}~\hlcyan{()} \\

\midrule

\multirow{5}{*}{\rotatebox[origin=c]{90}{\expttag{IA32\#260}}}
& \expttag{Inputs} & {move the 3rd element of the byte\_table into cl} \\
& \expttag{Outputs} & {mov cl, byte\_table+2} \\
& \exptlabel{Gen}{before} & \hlcyan{mov}~\hlcyan{ cl}~\hlcyan{,}~\hlcyan{ byte}~\hlcyan{\_}~\hlcyan{table}~\hlpink{[}~\hlpink{3} \\
& \exptlabel{Gen}{during} & \hlcyan{mov}~\hlcyan{ cl}~\hlcyan{,}~\hlcyan{ byte}~\hlcyan{\_}~\hlcyan{table}~\hlcyan{+}~\hlcyan{2} \\
& \exptlabel{Gen}{after} & \hlcyan{mov}~\hlcyan{ cl}~\hlcyan{,}~\hlcyan{ byte}~\hlcyan{\_}~\hlcyan{table}~\hlcyan{+}~\hlcyan{2} \\

\midrule

\multirow{5}{*}{\rotatebox[origin=c]{90}{\expttag{TLDR\#032}}}
& \expttag{Inputs} & {submit a job and request multiple nodes} \\
& \expttag{Outputs} & {sbatch --nodes=\{\{3\}\} \{\{path/to/job.sh\}\}} \\
& \exptlabel{Gen}{before} & \hlpink{q}~\hlcyan{batch}~\hlpink{ -}~\hlcyan{nodes}~\hlcyan{=\{\{}~\hlpink{2}~\hlcyan{\}\}}~\hlcyan{ \{\{}~\hlpink{script}~\hlcyan{/}~\hlcyan{to}~\hlcyan{/}~\hlpink{script}~\hlcyan{.}~\hlcyan{sh}~\hlcyan{\}\}} \\
& \exptlabel{Gen}{during} & \hlcyan{s}~\hlcyan{batch}~\hlcyan{ --}~\hlcyan{nodes}~\hlcyan{=\{\{}~\hlcyan{3}~\hlcyan{\}\}}~\hlcyan{ \{\{}~\hlcyan{path}~\hlcyan{/}~\hlcyan{to}~\hlcyan{/}~\hlcyan{job}~\hlcyan{.}~\hlcyan{sh}~\hlcyan{\}\}} \\
& \exptlabel{Gen}{after} & \hlcyan{s}~\hlcyan{batch}~\hlcyan{ --}~\hlcyan{nodes}~\hlcyan{=\{\{}~\hlcyan{3}~\hlcyan{\}\}}~\hlcyan{ \{\{}~\hlcyan{path}~\hlpink{\_}~\hlcyan{to}~\hlcyan{/}~\hlcyan{job}~\hlcyan{.}~\hlcyan{sh}~\hlcyan{\}\}} \\

\bottomrule

\end{tabular}

 }
\end{table}

The limitations of our approach \textsc{MINT} in successive LM repairs are outlined in the observations:
(1) In the \expttag{CoNaLa\#016} case, three incorrect tokens are processed through model repair and two of them are corrected. \textsc{MINT} is not able to solve all failures, indicating that there are opportunities for improving its effectiveness further;
(2) In the \expttag{IA32\#260} case, the tokens that were incorrectly predicted before model repair are successfully corrected. This is a positive case, showing the effectiveness of \textsc{MINT} remains unaffected in successive model repairs;
(3) In the \expttag{TLDR\#032} case, although all incorrect tokens are corrected, one token is incorrectly predicted after model repair, which was correctly predicted before and during the repair. It means, LM repair may affect the model's ability to predict the next token.

\subsection{Generalization and Specificity}

\begin{figure}[htbp]
    \centering
    \includegraphics[width=1.0\linewidth]{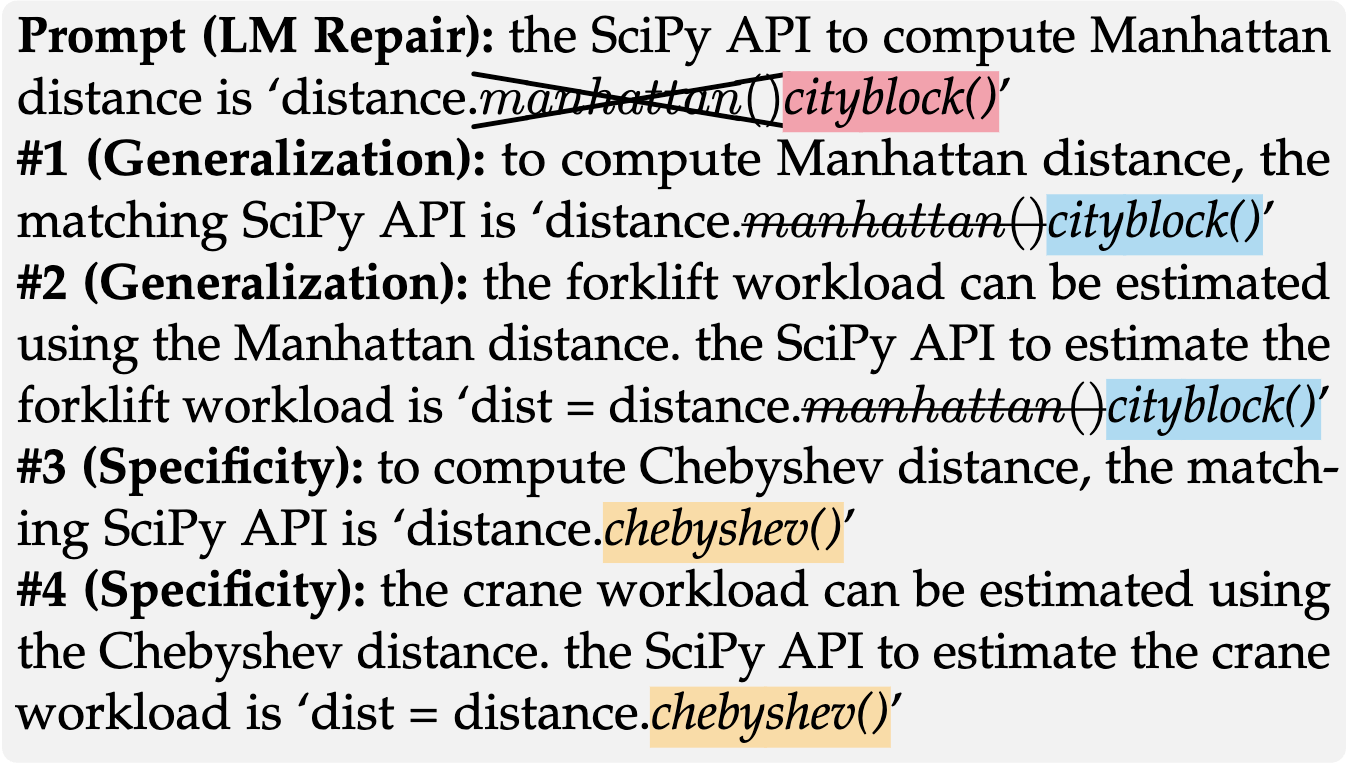}
    \caption{Demonstration of model repair for code generation.}
    \label{fig:example}
\end{figure}

To demonstrate how MINT handles the balance between the generalization and the specificity of LM repair, we consider an actual code generation case shown in \cref{fig:example}.
The cross symbol indicates the incorrectly predicted tokens by the code LM, namely the model failure, and are replaced with red-colored tokens by model repair.
The tokens marked with the strike-through symbol were initially incorrect tokens. They are the related data so are also affected and become blue-colored tokens after model repair.
The tokens in orange color are unaffected since they are the unrelated data, and they remain unchanged after model repair.

For an invalid API invocation to the non-existent function \textit{distance.manhattan()}, incorrectly generated by the \textsc{CodeGen} model, MINT corrects it with the accurate information as shown in \textbf{\textit{Prompt}}. This involves ensuring that the model recognizes \textit{distance.cityblock()} as the correct API to invoke.
First, we probe the generalization and specificity of the data with completely different expressions.
As shown in \textbf{\textit{\#1}}, the model now correctly predicts \textit{cityblock()} as the API of computing the Manhattan distance, rather than the original incorrect prediction \textit{manhattan()}.
Meanwhile, its behavior of computing the Chebyshev distance remains unaffected, still correctly using \textit{chebyshev()}, as shown in \textbf{\textit{\#3}}.
Additionally, we probe the generalization and specificity with more complex prompts, not directly asking the API, but asking the model to complete a line of code, by providing extra information.
As shown in \textbf{\textit{\#2}}, the repaired model correctly generates \textit{cityblock()} to estimate the forklift workload, rather than the original incorrect API \textit{manhattan()}.
Meanwhile, the model behavior of correctly generates \textit{chebyshev()} to estimate the crane workload remains unchanged, as shown in \textbf{\textit{\#4}}.
Overall, \textsc{MINT} shows good generalization on the related data (\textit{\#1} and \textit{\#2}), and good specificity on the unrelated data (\textit{\#3} and \textit{\#4}).

\section{Related Work}
\label{sec:related_work}

Model repair has been a significant and ongoing research topic in software engineering, and there haven been a certain amount of work. However, they are mainly proposed for CNN models, and discriminative tasks. They are hard to be directly adopted to address failures of language models, and generative tasks~\cite{Sohn2019ArachneSR,Gao2022FairneuronID}.
The general pattern of dealing with the problem of model repair is similar, such as locating the neurons with causal analysis~\cite{Sun2022CausalityBasedNN} and updating neurons with adaptive methods~\cite{LiCalsi2023AdaptiveSR}.
In addition, \textsc{VeRe}~\cite{Ma2024VereVG} proposed a verification-guided approach to repair CNN models, and validated the usefulness in backdoor removal. However, it is still not directly applicable to repairing language models. A significant challenge with verification-based techniques is their high computational cost, which remains difficult to mitigate. Besides, the differences between language models and other models (CNN, RNN, etc) indicate other unexpected difficulties.
Our approach \textsc{MINT} converts the process of solving model failures into eliminating the semantic difference in latent space, and proposes updating model parameters at the neuron level to reduce the side effects to the abilities of language models.



Knowledge editing methods are similar in methodology since they are also based on the theory of knowledge neurons but are mainly proposed for solving knowledge-related tasks, instead of general next-token prediction~\cite{Zhang2024ACS}.
Meanwhile, their methods tend to require additional data, so hard to be ready-to-use solutions for repairing language models.
Depending on whether additional components are introduced, there are two types of methods: intrinsic editing and extrinsic editing. The former directly updates the existing parameters of the model by itself, while the latter introduces additional editable parameters along with the model.



The typical intrinsic techniques include ROME~\cite{Meng2022LocatingAE} and MEMIT~\cite{Meng2022MassEditingMI}.
They narrow down the critical model layers and compute gradients to update model parameters.
They are proposed for knowledge synonym replacement, and cannot be directly used in next-token predictions.
These techniques require specifying an entity trigger (which is one or more entity tokens), so the usable data are mainly knowledge triples, limiting their use to synonym replacement.
Besides, they require a large corpus for sampling the knowledge and building the initial state of model parameters, which affects their effectiveness in adapting to other tasks.
The derived methods introduce the consideration of the interplays between different functional modules and realize further improvements in the performance, such as PMET~\cite{Li2023PMETPM}.

In contrast, extrinsic techniques are straightforward but lack flexibility and scalability.
GRACE~\cite{Hartvigsen2022AgingWG} utilizes codebooks between model layers to modify hidden representations, where codebooks are the combination of a classifier and a memory.
MEND~\cite{Mitchell2021FastME} and SERAC~\cite{Mitchell2022MemoryBasedME} can be applied to encoder-only models, decoder-only models, and encoder-decoder models. The former introduces a hyper-network for gradient decomposition, while the latter builds a parallel model to store edited behaviors and uses a classifier and a memory to decide which neural model to use. Overall, they are not editing the model but instead, tracking and responding to the model's internal activity.
They function similarly to rule-based methods but are more compatible with language models, and as a result, they tend to exhibit the same drawbacks as rule-based methods. 

\section{Conclusion}
\label{sec:conclusion}

In this paper, we studied an emerging topic, which is 
repair of code LMs.
Focused on repairing language models in code generation tasks, we proposed a semantic-based neuron-level approach \textsc{MINT}.
Experimental results show that our approach is effective and efficient. In addition, it achieves a good balance between generalization and specificity.
Based on our analysis of the showcase of code generation tasks, \textsc{MINT} is particularly useful as a hot-fix technique for LMs.

In our future work, we will explore applicable scenarios of model repair, especially tasks requiring high-quality reasoning, such as program comprehension~\cite{Liu2023CodeEW,Chen2024ReasoningRB}. Meanwhile, we plan to seek further improvements, for example, incorporating multiple stages into one to reduce the computation cost and improve the performance.





\bibliographystyle{IEEEtran}
\bibliography{references}

\end{document}